\documentclass[sigconf,screen]{acmart}

\AtBeginDocument{%
  }

\copyrightyear{2026}
\acmYear{2026}
\setcopyright{cc}
\setcctype{by}
\acmConference[SIGIR '26]{Proceedings of the 49th International ACM SIGIR Conference on Research and Development in Information Retrieval}{July 20--24, 2026}{Melbourne, VIC, Australia}
\acmBooktitle{Proceedings of the 49th International ACM SIGIR Conference on Research and Development in Information Retrieval (SIGIR '26), July 20--24, 2026, Melbourne, VIC, Australia}
\acmDOI{10.1145/3805712.3808549}
\acmISBN{979-8-4007-2599-9/2026/07}

\usepackage{subcaption}
\usepackage{tabularx}
\usepackage{graphicx}

\usepackage{subcaption}
\usepackage{caption}
\usepackage{enumitem}
\usepackage{adjustbox}
\usepackage{graphicx}
\usepackage{hyperref}
\usepackage{tikz}
\usepackage{makecell}
\usepackage{booktabs} 
\usetikzlibrary{positioning, arrows.meta, shapes.geometric, calc}
\usepackage[table]{xcolor} 

\usepackage{booktabs} 
\usepackage{calc}     
\usepackage{amsmath}  

\newcommand{\header}[1]{\vspace{2mm}\noindent\textbf{#1}}

\begin{document}

\title{Hypencoder Revisited: Reproducibility and Analysis of Non-Linear Scoring for First-Stage Retrieval}


\author{Arne Eichholtz}
\orcid{0009-0007-4816-2243}
\authornote{Equal Contribution.}
\affiliation{%
  \institution{University of Amsterdam}
  \city{Amsterdam}
  \country{The Netherlands}}
\email{arne.eichholtz@student.uva.nl}

\author{Yongkang Li}
\orcid{0000-0001-6837-6184}
\authornotemark[1]
\authornote{Corresponding author.}
\affiliation{%
  \institution{University of Amsterdam}
  \city{Amsterdam}
  \country{The Netherlands}}
\email{y.li7@uva.nl}

\author{Jutte Vijverberg}
\orcid{0009-0001-5447-408X}
\authornotemark[1]
\affiliation{%
  \institution{University of Amsterdam}
  \city{Amsterdam}
  \country{The Netherlands}}
\email{jutte.vijverberg@student.uva.nl}

\author{Tobias Groot}
\orcid{0009-0002-2132-0727}
\authornotemark[1]
\affiliation{%
  \institution{University of Amsterdam}
  \city{Amsterdam}
  \country{The Netherlands}}
\email{tobias.groot@student.uva.nl}

\author{Mohammad Aliannejadi}
\orcid{0000-0002-9447-4172}
\affiliation{%
  \institution{University of Amsterdam}
  \city{Amsterdam}
  \country{The Netherlands}}
\email{m.aliannejadi@uva.nl}


\begin{abstract}
  
The Hypencoder, proposed by \citet{killingback2025hypencoder}, is a retrieval framework that replaces the fixed inner-product scoring function used in standard bi-encoders with a query-specific neural network (the $q$-net), whose weights are generated by a hypernetwork from the contextualized query embeddings. This design enables more expressive relevance estimation while preserving independent query and document encoding. In this work, we conduct a reproducibility study of the Hypencoder and extend the original analysis in three directions. Our reproduction confirms that the Hypencoder outperforms a similarly trained bi-encoder baseline on in-domain and out-of-domain benchmarks, and that the proposed efficient search algorithm substantially reduces query latency with minimal performance loss. On hard retrieval tasks, we find partial support: the Hypencoder outperforms the baseline on DL-Hard and FollowIR, but not on TREC TOT, where checkpoint incompatibility and fine-tuning sensitivity complicate full verification. Beyond reproduction, we investigate three extensions: (i)~integrating alternative pre-trained encoders into the Hypencoder framework, where we find that performance gains depend on the encoder and fine-tuning strategy; (ii)~comparing query latency against a Faiss-based bi-encoder pipeline, revealing that standard bi-encoder retrieval remains faster under both exhaustive and efficient search settings; and (iii)~evaluating adversarial robustness, where we find that the $q$-net's non-linear scoring does not provide a consistent robustness disadvantage over inner-product scoring. Our code is publicly available at \url{https://github.com/arneeichholtz/Hypencoder-reprod}.
\end{abstract}




\maketitle

\section{Introduction}

Neural retrieval models face a fundamental trade-off between expressiveness and efficiency. Bi-encoder models encode queries and documents independently, enabling efficient retrieval through approximate nearest neighbor~(ANN) search~\cite{douze2024faiss}, but are constrained to simple scoring functions such as the inner product~\cite{xiao2022retromae,karpukhin2020dense}. Cross-encoders jointly encode the query--document pair and can capture complex interactions, but their computational cost makes them impractical for first-stage retrieval over large corpora~\cite{guo2020deep,khattab2020colbert,SanthanamKSPZ22_ColBERTv2}.

\begin{figure*}[t]
    \centering
    \includegraphics[width=0.9\linewidth]{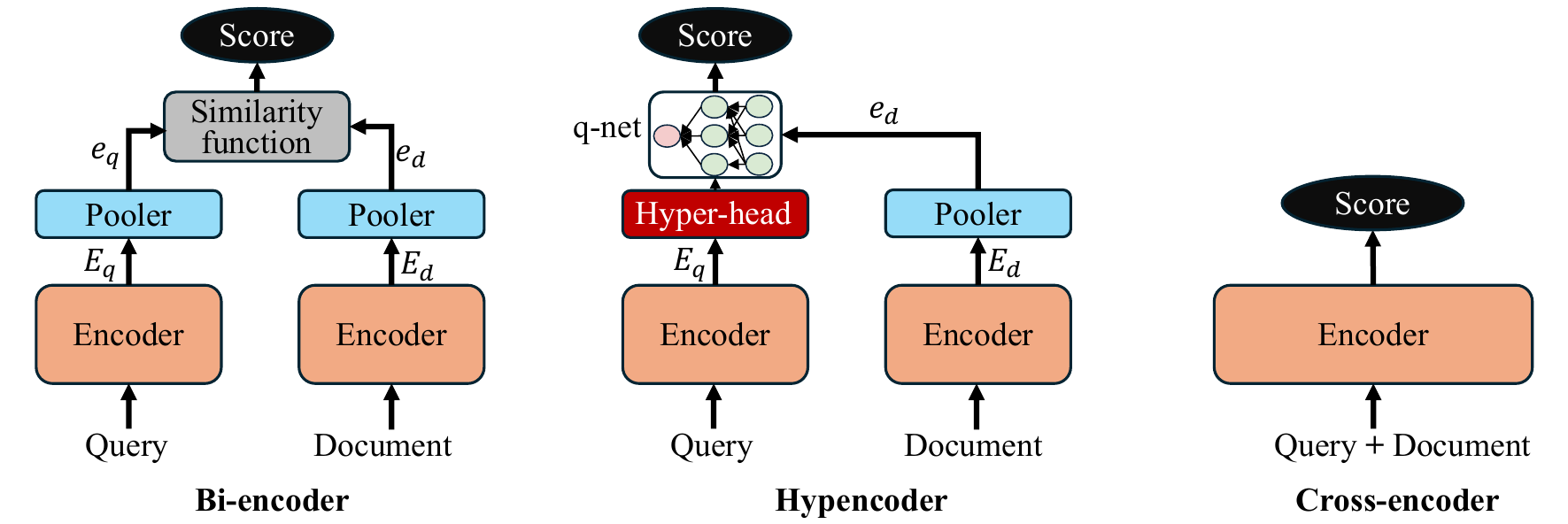}
    \caption{
Comparison of retrieval and reranking paradigms. While standard bi-encoders are limited by simple vector similarity (e.g., Inner Product) and cross-encoders suffer from high latency due to joint encoding, Hypencoder introduces a hybrid approach: using a hyperhead to dynamically generate a query-specific small neural network ($q$-net) that scores document representations independently and expressively. Figure adapted from \citet{killingback2025hypencoder}.
    }
   \label{fig:method_main}

\end{figure*}

\citet{killingback2025hypencoder} argue that this expressiveness gap is not merely an empirical shortcoming but a fundamental limitation: they provide a theoretical proof that, for any inner-product-based scoring function, there always exists a set of relevant documents that cannot be perfectly retrieved, regardless of the encoder used. To address this limitation, they propose the \emph{Hypencoder}, a new class of retrieval models that replaces the fixed inner product with a query-specific neural network as the scoring function. Specifically, contextualized query embeddings are processed by attention-based hypernetwork layers (the \emph{hyperhead}) to generate the weights and biases of a small feed-forward network (the \emph{$q$-net}). Document embeddings are then passed through the $q$-net to produce a scalar relevance score. This design preserves the efficiency of independent query and document encoding while enabling more expressive relevance estimation. To scale this approach to large corpora, the authors further propose a graph-based approximate search algorithm. An overview comparing the Hypencoder to existing retrieval paradigms is shown in Figure~\ref{fig:method_main}.

In this paper, we conduct a reproducibility study of the Hypencoder and evaluate the following claims from the original work:
\begin{enumerate}[label=\textbf{Claim \arabic*:}, leftmargin=*]
    \item The Hypencoder outperforms the baselines on in-domain 
          tasks.
    \item The Hypencoder outperforms the baselines on 
          out-of-domain tasks.
    \item The Hypencoder outperforms the baselines on hard 
          retrieval tasks (e.g., instruction-following retrieval).
    \item The efficient search algorithm improves query latency 
          while retaining retrieval performance.
\end{enumerate}
Our reproduction confirms Claims~1, 2, and~4. For Claim~3, we find partial support: the Hypencoder outperforms the bi-encoder baseline on DL-Hard and all three FollowIR subsets, but not on TREC TOT, where checkpoint incompatibility and sensitivity to fine-tuning hyperparameters complicate full verification. 

Beyond reproduction, we extend the analysis in three directions:

\header{Alternative encoders.}
Pre-training the Hypencoder from scratch is computationally expensive (six days on two A100 GPUs in the original work). We investigate whether the hypernetwork can be combined with existing pre-trained encoders through fine-tuning. We evaluate three alternative encoders and find that performance gains depend on the encoder and fine-tuning strategy used.

\header{Query latency analysis.}
The original paper does not compare the query latency against a standard bi-encoder retrieval pipeline. We fill this gap by benchmarking against a Faiss-based baseline across multiple datasets and corpus sizes, finding that Faiss-based retrieval is consistently faster.

\header{Adversarial robustness.}
We evaluate whether the $q$-net's non-linear scoring introduces new vulnerabilities to adversarial query perturbations and find that it does not yield a consistent robustness disadvantage compared to inner-product scoring.

Overall, the Hypencoder is a promising framework that achieves strong in-domain and out-of-domain performance with solid adversarial robustness. However, we identify two practical limitations: the substantial compute required to train or adapt the full Hypencoder pipeline beyond released checkpoints, and higher query latency compared to standard bi-encoder pipelines.

\section{Related Work} 

We review multi-stage neural information retrieval and situate Hypencoder in first-stage retrieval. We then summarize hypernetworks, the architectural basis of Hypencoder.

\subsection{Neural Information Retrieval}
Modern information retrieval systems typically adopt a multi-stage retrieve-then-rerank paradigm to balance efficiency and effectiveness~\cite{nogueira2019passage, wang2011cascade, lin2021pretrained}.

\header{First-stage retrieval.}
First-stage retrievers are generally categorized into sparse and dense approaches. Sparse methods, exemplified by BM25~\cite{RobertsonWJHG94_BM25}, rely on lexical matching. Recent advances like SPLADE~\cite{formal2021splade} bridge the semantic gap by performing learned term expansion while retaining the efficiency of inverted indices. 

In the realm of dense retrieval, the bi-encoder architecture has become the de facto standard~\cite{karpukhin2020dense}. By encoding queries and documents independently into separate embeddings, bi-encoders allow document representations to be pre-computed and indexed for Maximum Inner Product Search (MIPS). Significant progress has been made in optimizing bi-encoders through contrastive learning (e.g., Contriever~\cite{izacard2022unsupervised}), negative sampling strategies (e.g., ANCE~\cite{xiong2020approximate}, DRAGON~\cite{LinALOLMY023}), and distillation (e.g., TAS-B~\cite{hofstatter2021efficiently}).
More recently, LLM-based dense retrievers such as repLLaMA~\cite{MaWYWL24_repllama} and E5-Mistral \cite{WangYHYMW24_e5} have achieved strong performance by leveraging large language models as backbone encoders. However, all standard bi-encoders share a common limitation: they rely on a simple, fixed similarity function (typically dot product or cosine similarity), which compresses the query's semantic intent into a single geometric point, potentially creating a bottleneck for complex matching patterns.
To mitigate this bottleneck, late interaction models like ColBERT~\cite{khattab2020colbert, SanthanamKSPZ22_ColBERTv2} decompose texts into token-level representations and compute relevance through a MaxSim operation. While this approaches the expressiveness of cross-encoders, it incurs a significantly higher storage cost compared to single-vector dense retrieval.\looseness=-1

\header{Second-stage reranking.}
The second stage typically employs cross-encoders, which jointly encode query-document pairs for fine-grained relevance modeling~\cite{nogueira2019passage}.  Approaches include pointwise methods such as  MonoBERT~\cite{nogueira2019passage} and MonoT5 \cite{NogueiraJPL20_monoT5}, which predict absolute relevance scores;  pairwise methods like DuoBERT~\cite{nogueira2019multi}, which model relative preferences between document pairs;  and listwise methods like RankGPT~\cite{sun2023chatgpt} and rankLLaMA~\cite{MaWYWL24_repllama}, which optimize over entire ranked lists. Knowledge distillation from cross-encoders to bi-encoders has also proven effective for improving first-stage retrieval~\cite{hofstatter2021efficiently, lin2020distilling, SanthanamKSPZ22_ColBERTv2}.

\header{Positioning hypencoder.}
Hypencoder \cite{killingback2025hypencoder} targets the first-stage retrieval setting but departs from standard dense retrieval in a key way: instead of representing a query as a single embedding vector and using a \emph{fixed} similarity function, it produces a \emph{query-specific} relevance function. Concretely, Hypencoder generates a small scoring network (the $q$-net) conditioned on the query, and uses this network to map a document representation to a relevance score. This design occupies an intermediate position between bi-encoders, which have limited expressiveness due to fixed similarity functions, and cross-encoders, which are expressive but too expensive for full-corpus retrieval. In this sense, Hypencoder aims to increase first-stage scoring expressiveness while retaining scalability by operating on pre-computed document representations and employing approximate search procedures \cite{killingback2025hypencoder}.

\subsection{Hypernetworks}

Hypernetworks,  introduced by Ha et al.~\cite{ha2016hypernetworks}, represent a class of neural architectures where one network (the \textit{hypernetwork}) generates the weights for another network (the \textit{target} or \textit{main} network).
The core intuition behind hypernetworks is adaptability. Instead of learning a single set of static weights for a task, the model learns a function that maps context (e.g., a task description or input embedding) to a specific weight configuration. This paradigm has been successfully applied across various domains~\cite{chauhan2024brief} such as neural architecture search~\cite{zhang2018graph,brock2018smash}, continual learning~\cite{Oswald2020Continual},   federated learning~\cite{shamsian2021personalized}, and  multi-task learning~\cite{navon2021learning}

In standard applications, hypernetworks are typically simple and small networks that generate weights for a larger main network to save parameters. However, the \textit{Hypencoder} reverses this architectural convention for the sake of retrieval efficiency. Here, the hypernetwork is a large, powerful transformer (the query encoder), while the target network ($q$-net) is a small Multi-Layer Perceptron (MLP). To the best of our knowledge, Hypencoder represents the first successful application of hypernetworks to first-stage retrieval, using the hypernetwork to encode the query semantics into $q$-net parameters, effectively generating a query-specific function for scoring documents in the embedding space.  
 
\section{Reproducibility Methodology}
In this section, we outline the core methodology of the Hypencoder framework \cite{killingback2025hypencoder}. We first formalize the retrieval problem and then describe the unified Hypencoder architecture, which integrates the hypernetwork mechanism for query-specific scoring. Finally, we detail the efficient graph-based search procedure.

\subsection{Problem Formulation}
Let $\mathcal{D} = \{d_1, \dots, d_N\}$ denote a corpus and $q$ a query. In standard dense retrieval, documents and queries are encoded independently. Let $E_q \in \mathbb{R}^{n \times h}$ and $E_d \in \mathbb{R}^{m \times h}$ denote the contextualized token-level representations for the query and document, respectively. 
Existing bi-encoder models typically pool these representations into fixed-size vectors $e_q, e_d \in \mathbb{R}^{h}$ and estimate the relevance score $s(q,d)$ using a fixed similarity function $\psi$:
\begin{equation}
    s(q,d) = \psi(e_q, e_d)~,
\end{equation}
where $\psi$ is commonly the inner product or cosine similarity. The Hypencoder replaces the fixed $\psi$ and the pooled query vector $e_q$ with a learned, query-dependent neural network.

\subsection{Hypencoder Architecture}
The Hypencoder architecture consists of three main components: a query encoder with a hyperhead, a document encoder, and the generated query-specific network, the $q$-net.

\header{Query encoder and hyperhead.}
Unlike standard bi-encoders, Hypencoder utilizes the full sequence of non-pad query embeddings $E_q$ (as defined above) to generate the parameters $\theta_q$ for the $q$-net. This generation process is handled by the \textit{hyperhead}.

To generate a specific weight matrix $W \in \mathbb{R}^{r \times t}$ for a target layer in the $q$-net, the hyperhead applies the following transformation pipeline:

\begin{itemize}
    \item \textbf{Input expansion:} The query embeddings $E_q$ are concatenated with a column of ones to jointly model weights and biases. 
    \item \textbf{Attention mechanism:} This expanded input is projected into keys and values. A set of learned query vectors $Q$ then performs scaled dot-product attention over them to produce latent representations.
    \item \textbf{Non-Linear projection:} The attention output undergoes a ReLU activation and Layer Normalization, followed by a linear projection to match the target dimension $r \times t$.
    \item \textbf{Base weight addition:} Finally, the projected update is added to a learnable, query-independent base matrix $\theta_{base}$. This residual formulation allows the model to refine a universal matching pattern with query-specific adjustments.
\end{itemize}

\header{Document encoder.}
Consistent with the problem formulation, the document encoder outputs token representations $E_d$. To maintain storage efficiency, $E_d$ is pooled using the \texttt{[CLS]} token into a single vector $e_d \in \mathbb{R}^{h}$.

\header{The $q$-net scoring function.}
The parameters generated by the hyperhead are assembled into the $q$-net. The pooled document vector $e_d$ serves as the input to this network:
\begin{equation}
    s(q, d) = q\text{-net}(e_d; \theta_q)~.
\end{equation}
In the original work, the $q$-net is a lightweight feed-forward network where each intermediate layer consists of a linear transformation (using the generated weights), followed by a ReLU activation, Layer Normalization, and a residual connection. The final layer is a linear projection to a scalar relevance score, without a residual connection or non-linearity.

\subsection{Efficient Retrieval with Hypencoder}
Efficient retrieval is crucial for scaling to large corpora. Standard ANN search methods rely on the assumption that the scoring function correlates with the geometric structure of the embedding space: documents that are close in vector space (e.g., via inner product or cosine similarity) typically yield similar relevance scores for a given query.
However, the Hypencoder’s $q$-net is a learned, non-linear scoring function. Consequently, local geometric proximity does not guarantee similar relevance scores, as small changes in the document vector can trigger significant changes in the $q$-net output. Because standard ANN indexing cannot be directly applied, \citet{killingback2025hypencoder} propose a greedy search algorithm over a document-to-document graph.\looseness=-1

\header{Graph construction.}
In the offline indexing stage, a neighbor graph is constructed where each document acts as a node. Edges are created by connecting each node to its 100 nearest neighbors based on the  Euclidean $\ell_2$  distance between document embeddings. This graph structure enables efficient traversal without relying on inner-product assumptions.

\header{Inference procedure.}
During retrieval, the algorithm utilizes the generated query-specific $q$-net to navigate this graph. The search begins with a set of randomly sampled candidate documents $\tilde{C}$. The process then iterates through the following steps until a termination condition is met:

\begin{itemize}
    \item \textbf{Scoring and selection:} All candidates in the current pool $\tilde{C}$ are scored using the $q$-net. From this pool, the top $nCandidates$ with the highest scores are selected for processing.
    \item \textbf{Result update:} The algorithm updates the global top-$k$ results set, $T$, using the selected candidates. If a selected candidate has a higher score than the lowest-scoring document in $T$, it is inserted into $T$. 
    \item \textbf{Expansion:} The search frontier is expanded by retrieving the unvisited neighbors of the selected candidates. These neighbors form the candidate pool $\tilde{C}$ for the next iteration.
\end{itemize}

The iterative search terminates when one of the following occurs: (i) the candidate pool becomes empty; (ii) the highest-scoring candidate in the current pool scores below the lowest-scoring element in $T$, indicating that no candidate can improve the current result (early stopping); or (iii) a maximum number of iterations ($\textit{maxIter}$) is reached.
The runtime complexity is approximately $\mathcal{O}(|\tilde{C}| + nCandidates \cdot \textit{maxIter})$, effectively decoupling query latency from the total corpus size.

\header{Search configurations.}
The performance of this algorithm depends on three hyperparameters: the initial candidate size $|\tilde{C}|$, the expansion beam size $nCandidates$, and the iteration limit $\textit{maxIter}$. Based on empirical analysis on the MS MARCO corpus, the original authors define two standard configurations, which we adopt:

\begin{itemize}
    \item \textbf{Efficient 1:} A speed-oriented setting with $|\tilde{C}|=10{,}000$, $nCandidates = 64$, and $maxIter=16$.
    \item \textbf{Efficient 2:} A quality-oriented setting with $|\tilde{C}|=100{,}000$, $nCandidates = 328$, and $maxIter=20$.
\end{itemize}
\section{Experimental Setup}
\label{sec:setup}
In this section, we describe the experimental setup for the reproducibility study. We detail the datasets, models, training procedures, evaluation metrics, and implementation specifics. The setup for the extension experiments is described separately in Section~\ref{sec:extension_setup}.

\subsection{Datasets}
\label{sec:datasets}

\header{In-Domain evaluation.}
We evaluate on the MS MARCO Passage Ranking dataset~\cite{nguyen2016ms}, which contains approximately 7k queries with shallow relevance labels over a corpus of 8.8M passages. We also evaluate on the TREC Deep Learning 2019~\cite{craswell2020overview} and 2020~\cite{craswell2021overviewtrec2020deep} passage datasets, which share the same corpus but provide deeper annotations over a smaller set of queries (97 queries combined).

\header{Out-of-Domain evaluation.}
The original paper evaluates out-of-domain performance on five selected BEIR datasets. To provide a more comprehensive assessment of out-of-domain generalization, we extend this evaluation to all 13 publicly available datasets in the BEIR benchmark~\cite{thakur2021beir}, covering question answering (NQ~\cite{KwiatkowskiPRCP19_NQ}, HotpotQA~\cite{Yang0ZBCSM18_HotpotQA}, FiQA~\cite{maia201818}), biomedical and scientific retrieval (TREC-COVID~\cite{roberts2021searching}, NFCorpus~\cite{boteva2016full}, SciFact~\cite{WaddenLLWZCH20_scifact}, SCIDOCS~\cite{CohanFBDW20_scidocs}), fact verification (FEVER~\cite{ThorneVCM18_FEVER}, Climate-FEVER~\cite{abs-2012-00614_CLIMATE-FEVER}), entity retrieval (DBpedia~\cite{hasibi2017dbpedia}), argument retrieval (Touch\'{e}-2020~\cite{bondarenko2020overview}, ArguAna~\cite{WachsmuthSS18_arguana}), and duplicate question detection (Quora~\cite{quora2017dataset}). We report the original five datasets separately to enable direct comparison with the reported results.

\header{Hard retrieval tasks.}
We consider three datasets for evaluating performance on more challenging retrieval tasks. First, the TREC Tip-of-the-Tongue (TOT) Track 2023~\cite{arguello2023overview}, which features verbose, multi-facet queries; we use the 150-query development set, as the test set relevance labels were not publicly available at the time of the original work. Second, FollowIR~\cite{weller2025followir}, which evaluates instruction-following capability in retrieval across three subsets (Robust~'04, News~'21, and Core~'17). Third, a subset of TREC DL-Hard~\cite{mackie2021deeplearningdlhardannotated}, restricted to queries with TREC-provided relevance judgments. This restriction follows the original authors' decision, as they found that queries labeled by the DL-Hard authors had substantially fewer judged documents in the top 10 (approximately 15\%) compared to those with TREC labels (approximately 93\%), making evaluation metrics unreliable for the former.

\header{Fine-Tuning data for hard tasks.}
For TREC TOT, we use the Reddit-sourced training data introduced by \citet{bhargav2022s}. For FollowIR, we use MS MARCO with Instructions~\cite{weller2024promptriever}. No additional fine-tuning is performed for TREC DL-Hard, as the base Hypencoder model is already trained on MS MARCO.

\subsection{Models}\label{sec:models}

\citet{killingback2025hypencoder} introduce two retrieval models: Hypencoder and BE-Base. The Hypencoder uses a \textbf{shared} BERT base (uncased) encoder for both queries and documents, with a 6-layer $q$-net (reported as the optimal configuration). BE-Base is a standard bi-encoder baseline trained under the same conditions as the Hypencoder but \textbf{using separate query and document encoders} and an inner product scoring function, enabling a controlled comparison that isolates the effect of the hypernetwork architecture.

We use the publicly available Hypencoder checkpoint on Hugging Face.\footnote{\url{https://huggingface.co/jfkback/hypencoder.6_layer}} The BE-Base checkpoint, including both the MS MARCO-trained version and the hard-task fine-tuned versions, were obtained through direct communication with the authors.

\subsection{Training}
\label{sec:training}

For the in-domain and out-of-domain evaluation, we use the provided model checkpoints directly and do not perform any additional training. For the hard retrieval tasks, we initialize from the MS MARCO-trained Hypencoder checkpoint and perform task-specific fine-tuning. Following the original paper, we use AdamW with a learning rate of $8\times10^{-6}$, a linear scheduler with a warm-up ratio of 0.1, and cross-entropy loss only (Margin MSE is not used for hard task fine-tuning). For TREC TOT, we fine-tune for 25 epochs ($\sim$3.3k steps) with a batch size of 96. For FollowIR, we fine-tune for 1 epoch ($\sim$10k steps) with the same batch size. Each training example consists of a query, a positive document, and a hard negative document, with a maximum sequence length of 512 tokens for both queries and documents. No additional fine-tuning is performed for TREC DL-Hard. The fine-tuned BE-Base checkpoints for the hard tasks were obtained from the authors. The exact configuration files used for our fine-tuning runs are available in our GitHub repository.

\subsection{Evaluation Metrics}
\label{sec:metrics}

\header{Retrieval quality.}
We adopt the same evaluation metrics as the original paper. For the in-domain tasks, we report \textbf{nDCG@10}, \textbf{MRR}, and \textbf{R@1000} on the TREC Deep Learning 2019 and 2020 datasets, and \textbf{MRR@10} and \textbf{R@1000} on the MS MARCO Dev dataset. For the out-of-domain tasks, performance is evaluated using \textbf{nDCG@10}.
For the hard retrieval tasks, we follow the original evaluation protocols. On TREC DL-Hard, we report \textbf{nDCG@10}, \textbf{MRR}, and \textbf{R@1000}. For TREC TOT, we report \textbf{nDCG@10}, \textbf{MRR}, and \textbf{nDCG@1000}. For FollowIR, we report \textbf{AP} and \textbf{p-MRR} on Robust~'04 and Core~'17, and \textbf{nDCG@5} and \textbf{p-MRR} on News~'21, following the per-subset metrics used in the original paper. The p-MRR metric~\cite{weller2024followirevaluatingteachinginformation} measures how a model's document ranking changes in response to modified retrieval instructions: a value of $+100$ indicates a perfect re-ranking, 0 indicates no change, and $-100$ indicates the opposite of the desired change.

\header{Query latency.}
We report query latency as the average wall-clock time per query to execute the retrieval function. For the exhaustive Hypencoder, this includes $q$-net generation and batch-wise scoring over the full corpus. For the efficient search variants, this includes $q$-net generation followed by the graph-based search algorithm; the offline neighbor-graph construction cost is reported separately (Figure~\ref{fig:graph_construction_time}). For the Faiss-based bi-encoder baseline, we use an \texttt{IndexFlatIP} index from Faiss~\cite{johnson2019billion} and report amortized end-to-end latency: $(T_{\text{build}} + T_{\text{search}}) / N_{\text{queries}}$, where $T_{\text{build}}$ is the one-time index construction and $T_{\text{search}}$ the total search time over all queries. Faiss does not natively support BF16 precision; we therefore use FP32 precision for all Faiss experiments.

\subsection{Implementation Details}
\label{sec:implementation}

The original codebase is well-documented and publicly available.\footnote{\url{https://github.com/jfkback/hypencoder-paper}} It includes replication commands for the in-domain and out-of-domain experiments, which we used directly. The provided commands ran successfully without any code modifications.

For the hard retrieval tasks, additional implementation steps were necessary. First, we performed data preprocessing to prepare the datasets for the original codebase's expected format; the corresponding scripts are provided in our repository. Second, we implemented the fine-tuning pipeline as described in Section~\ref{sec:training}. Third, since the original codebase did not include an implementation of the p-MRR metric, we adopted the implementation from MTEB.\footnote{\url{https://github.com/embeddings-benchmark/mteb}} All experiments were conducted on NVIDIA H100 GPUs.

\subsection{Extension Experiment Setup}
\label{sec:extension_setup}

This section describes the additional setup required for the extension experiments beyond the reproducibility study. Unless otherwise noted, the evaluation metrics and base implementation are the same as described above.

\header{Alternative encoders.}
To investigate the generalizability of the Hypencoder framework across different encoder backbones, we replace the shared BERT encoder with three alternative pre-trained encoders: TAS-B\footnote{\url{https://huggingface.co/sebastian-hofstaetter/distilbert-dot-tas_b-b256-msmarco}}~\cite{hofstatter2021efficiently}, RetroMAE\footnote{\url{https://huggingface.co/Shitao/RetroMAE_MSMARCO_finetune}}~\cite{xiao2022retromae}, and Contriever\footnote{\url{https://huggingface.co/facebook/contriever-msmarco}}~\cite{izacard2022unsupervised}. Concretely, we initialize the Hypencoder's encoder with each alternative model's pre-trained weights, attach the hyperhead, and fine-tune on the MS MARCO training dataset provided by \citet{killingback2025hypencoder}.\footnote{\url{https://huggingface.co/datasets/jfkback/hypencoder-msmarco-training-dataset}} 
We consider two fine-tuning settings: (1)~fine-tuning only the hyperhead while keeping the encoder frozen (learning rate $1\times10^{-3}$), and (2)~jointly fine-tuning both the encoder and the hyperhead (learning rate $1\times10^{-5}$). The higher learning rate in the frozen setting reflects the small number of randomly initialized hyperhead parameters, whereas the end-to-end setting uses a standard transformer fine-tuning rate. Both settings are trained for three epochs, following standard practice for dense retrieval training on MS MARCO~\cite{WangYHJYJMW23_SimLM}. We note that the original Hypencoder was trained from a raw BERT checkpoint for 800k steps; our setup instead leverages encoders that are already fine-tuned for retrieval, which reduces the required training budget.

\header{Adversarial robustness.}
To evaluate robustness against adversarial query perturbations, we apply the query variation generation framework introduced by \citet{penha2022evaluating} to the in-domain datasets. We use the same TAS-B checkpoint as in the alternative encoder experiments as a baseline for comparison.

The results are organized into two parts: Section~\ref{sec:reprod_results} evaluates the four claims from \citet{killingback2025hypencoder}, while Section~\ref{sec:ext_results} presents our extension analyses.

\section{Reproduction Results}\label{sec:reprod_results}

We present our results organized by the four claims from \citet{killingback2025hypencoder}. For each claim, we first assess reproducibility by comparing our results to the original paper (indicated by~(O) in the tables, with reproduced results indicated by~(R)), then evaluate the claim by comparing the Hypencoder against the baselines.

\begin{table}[t]
\small
\caption{Evaluation on in-domain tasks, including Original (O) and Reproduced (R) values for BE-Base and Hypencoder, and results for six baselines. Values are taken from the cited sources unless otherwise stated; (–) indicates not reported.}
\label{tab:rep_indomain}
\centering
\setlength{\tabcolsep}{4pt}
\renewcommand\arraystretch{1}
\begin{tabular}{lccc|cc}
\toprule
& \multicolumn{3}{c}{\textbf{TREC-DL '19 \& '20}} & \multicolumn{2}{c}{\textbf{MS MARCO Dev}} \\
\textbf{Model} & nDCG@10 & MRR & R@1000 & MRR@10 & R@1000 \\
\midrule
BM25 & 0.491 & 0.679 & 0.735 & 0.184 & 0.853 \\
TAS-B & 0.700 & 0.863 & 0.861 & 0.344 & 0.978 \\
Contriever & 0.615 & -- & -- & 0.341 & 0.980 \\
RetroMAE & 0.694 & -- & -- & \textbf{0.416} & \underline{0.988} \\
repLLaMA & 0.731 & -- & -- & \underline{0.412} & \textbf{0.994} \\
ColBertv2 & \textbf{0.749} & -- & -- & 0.397 & 0.984 \\
\midrule
BE-Base (O) & 0.713 & 0.855 & \underline{0.868} & 0.359 & 0.980 \\
BE-Base (R) & 0.712 & \underline{0.950} & 0.796 & 0.371 & 0.980 \\
\midrule
Hypencoder (O)  & \underline{0.736} & 0.885 & \textbf{0.871} & 0.386 & 0.981 \\
Hypencoder (R) & 0.735 & \textbf{0.966} & 0.813 & 0.387 & 0.982 \\
\bottomrule
\end{tabular}
\end{table}

\begin{table*}[t]
\caption{Evaluation on out-of-domain tasks in nDCG@10. Original (O) rows report values from the original paper; Reproduced (R) rows report our evaluation using the released checkpoints. (--) indicates the result was not reported in the original paper.}
\centering
\small
\setlength{\tabcolsep}{0.6pt}
\renewcommand\arraystretch{1}
\begin{tabular}{lccccc | cccccccc}
\toprule
\textbf{Model} & \textbf{TREC-COVID} & \textbf{FiQA} & \textbf{NFCorpus} & \textbf{DBPedia} & \textbf{Touch\'{e}-2020} & \textbf{NQ} & \textbf{HotpotQA} & \textbf{ArguAna} & \textbf{Quora} & \textbf{SCIDOCS} & \textbf{FEVER} & \textbf{Climate-FEVER} & \textbf{SciFact} \\
\midrule
BM25 & 0.656 & 0.236 & 0.325 & 0.177 & \textbf{0.367} & 0.329 & 0.503 & 0.315 & 0.789 & 0.158 & 0.753 & 0.213 & 0.665 \\
TAS-B & 0.481 & 0.300 & 0.319 & 0.384 & 0.162 & 0.463 & 0.584 & 0.429 & 0.835 & 0.149 & 0.700 & 0.228 & 0.643 \\
Contriever  & 0.596 & 0.329 & 0.328 & 0.413 & 0.230 & 0.498 & 0.638 & 0.446 & \underline{0.865} & \underline{0.165} & 0.758 & 0.237 & 0.677 \\
RetroMAE  & \underline{0.772} & 0.316 & 0.308 & 0.390 & 0.237 & 0.518 & 0.635 & 0.433 & 0.847 & 0.150 & 0.774 & 0.232 & 0.653 \\
repLLaMA  & \textbf{0.847} & \textbf{0.458} & \textbf{0.378} & \underline{0.437} & 0.305 & \textbf{0.624} & \textbf{0.685} & \textbf{0.486} & \textbf{0.868} & \textbf{0.181} & \textbf{0.834} & \textbf{0.310} & \textbf{0.756} \\
ColBERTv2 & 0.738 & \underline{0.356} & \underline{0.338} & \textbf{0.446} & 0.263 & \underline{0.562} & \underline{0.667} & \underline{0.463} & 0.852 & 0.154 & \underline{0.785} & 0.176 & \underline{0.693} \\
\midrule
BE-Base (O) & 0.651 & 0.309 & 0.327 & 0.405 & 0.240 & -- & -- & -- & -- & -- & -- & -- & --  \\
BE-Base (R) & 0.603 & 0.309 & 0.321 & 0.393 & \underline{0.303} & 0.439 & 0.556 & 0.336 & 0.832 & 0.137 & 0.705 & \underline{0.254} & 0.597 \\
\midrule
Hypencoder (O) & 0.688 & 0.314 & 0.324 & 0.419 & 0.258 & -- & -- & -- & -- & -- & -- & -- & --  \\
Hypencoder (R) & 0.698 & 0.314 & 0.324 & 0.419 & 0.258 & 0.529 & 0.589 & 0.321 & 0.805 & 0.147 & 0.707 & 0.222 & 0.623 \\

\bottomrule
\end{tabular} 
\label{tab:rep_outdomain}
\end{table*}

\subsection{Claim 1: Hypencoder Outperforms the Baselines on In-Domain Tasks}\label{sec:claim1}
Table~\ref{tab:rep_indomain} presents the in-domain results. The original paper includes a large set of baselines and reference models. For our reproduction, we select a representative subset: BM25 as a sparse baseline, TAS-B as a distillation-trained dense retriever, and BE-Base as the controlled bi-encoder baseline from the original work. To broaden the comparison, we additionally include Contriever and RetroMAE as two widely adopted dense retrievers of comparable scale. Following the original paper's 
distinction, we also report two reference models that differ in architecture or scale (repLLaMA: 7B parameters; ColBERTv2: multi-vector), which are not directly comparable but provide useful context.

\header{Reproducibility.}
On MS MARCO Dev, the reproduced Hypencoder results closely match the original (MRR@10: 0.387 vs 0.386). On the combined TREC DL~'19~\&~'20, nDCG@10 is reproduced accurately (0.735 vs 0.736), but we observe deviations in MRR (0.966 vs 0.885) and R@1000 (0.813 vs 0.871). This pattern of elevated MRR and reduced R@1000 also appears for BE-Base and recurs in our DL-Hard results (Table~\ref{tab:hardtasks}). The systematic nature of this shift---consistent across both models and multiple evaluation sets---suggests a shared pipeline-level factor rather than a model-specific issue. Since nDCG@10 reproduces accurately and the relative ordering between models is preserved, we base our claim evaluation primarily on this metric.\looseness=-1

\header{Claim evaluation.}
Against the original baselines, the Hypencoder achieves the highest nDCG@10 on the combined TREC datasets (0.735), outperforming BM25~(0.491), TAS-B~(0.700), and BE-Base~(0.712). On MS MARCO Dev, the Hypencoder's MRR@10~(0.387) likewise surpasses BE-Base~(0.371) and TAS-B~(0.344). These results are consistent with the original findings. We therefore \textbf{confirm Claim~1} within the scope of the original baselines.

When extending the comparison to additional baselines, we observe that RetroMAE achieves a higher MRR@10 on MS MARCO Dev~(0.416 vs 0.387), indicating that the Hypencoder's advantage is not universal across all comparably-scaled dense retrievers. The reference models also outperform the Hypencoder on MS MARCO Dev but not consistently on the TREC datasets (e.g., the Hypencoder's nDCG@10 of 0.735 exceeds repLLaMA's 0.731), suggesting that the Hypencoder is competitive even with substantially larger models on deeper-annotated benchmarks.

\begin{table*}[t]
\caption{Evaluation on hard tasks. Row labels indicate the result source: values reported in the original paper (Orig.\ Reported), our evaluation of released checkpoints (Orig.\ Ckpts),  the base model from Section~\ref{sec:claim1} without task-specific fine-tuning (No FT), and our fine-tuned reproduction (FT Ours). The original Hypencoder fine-tuned checkpoints were incompatible with the released codebase and are therefore omitted. (–) indicates not applicable.}
\label{tab:hardtasks}
\centering
\small
\setlength{\tabcolsep}{3.5pt}
\renewcommand\arraystretch{1}
\begin{tabular}{l ccc | ccc | cc | cc | cc}
\toprule
& \multicolumn{3}{c|}{\textbf{TREC DL-HARD}} 
& \multicolumn{3}{c|}{\textbf{TREC TOT DEV}} 
& \multicolumn{2}{c|}{\textbf{FollowIR Robust '04}} 
& \multicolumn{2}{c|}{\textbf{FollowIR News '21}} 
& \multicolumn{2}{c}{\textbf{FollowIR Core '17}} \\
\cmidrule(lr){2-4} \cmidrule(lr){5-7} \cmidrule(lr){8-9} \cmidrule(lr){10-11} \cmidrule(lr){12-13}
\textbf{Model} 
& nDCG@10 & MRR & R@1000
& nDCG@10 & MRR & nDCG@1000
& AP & p-MRR
& nDCG@5 & p-MRR
& AP & p-MRR \\
\midrule
\multicolumn{13}{l}{\textit{BE-Base}} \\
Orig. Reported  & 0.607 & 0.864 & 0.805 & 0.121 & 0.110 & 0.179 & 0.207 & -3.7 & 0.239 & -1.1 & 0.178 & -7.7 \\
Orig. Ckpts     & 0.601 & 0.931 & 0.751 & 0.121 & 0.111 & 0.180 & 0.238 & -1.9 & 0.415 &  0.0 & 0.266 &  4.0 \\
No FT           & 0.601 & 0.931 & 0.751 & 0.022 & 0.016 & 0.207 & 0.220 & -4.6 & 0.404 & -0.1 & 0.260 &  3.6 \\
FT (Ours)       & ---   & ---   & ---   & 0.077 & 0.071 & 0.320 & 0.232 & -1.7 & 0.399 & -0.4 & 0.257 &  1.9 \\
\midrule
\multicolumn{13}{l}{\textit{Hypencoder}} \\
Orig. Reported  & 0.630 & 0.887 & 0.798 & 0.134 & 0.125 & 0.182 & 0.212 & -3.5 & 0.272 &  2.0 & 0.193 & -11.8 \\
No FT           & 0.630 & 0.958 & 0.759 & 0.028 & 0.026 & 0.055 & 0.256 & -2.6 & 0.395 &  0.3 & 0.280 &  0.6 \\
FT (Ours)       & ---   & ---   & ---   & 0.068 & 0.066 & 0.113 & 0.262 & -3.5 & 0.419 & 0.1 & 0.286 &  0.3 \\
\bottomrule
\end{tabular}
\end{table*}

\subsection{Claim 2: Hypencoder Outperforms the Baselines on Out-of-Domain Tasks}
\sloppy{Table~\ref{tab:rep_outdomain} presents the out-of-domain results in} nDCG@10. The original paper reports results on five BEIR datasets with TAS-B, CL-DRD, and BE-Base as baselines. We reproduce the same five datasets using our selected baseline set (Section~\ref{sec:claim1}) and extend the evaluation to all 13 publicly available BEIR datasets to provide a more comprehensive assessment of out-of-domain generalization.

\header{Reproducibility.}
On the five original datasets, the reproduced Hypencoder scores match exactly or near-exactly (e.g., FiQA: 0.314 vs 0.314; DBPedia: 0.419 vs 0.419), confirming the reliability of our evaluation setup.

\header{Claim evaluation.}
On the five original datasets, the Hypencoder outperforms TAS-B on all five and BE-Base on four out of five, consistent with the original findings. We therefore \textbf{confirm Claim~2} within the original evaluation scope.

To assess the generality of the out-of-domain advantage, we extend the evaluation to all 13 publicly available BEIR datasets. On the eight additional datasets, the Hypencoder's advantage diminishes: it outperforms TAS-B on only three and BE-Base on four datasets. Contriever and RetroMAE also outperform the Hypencoder on the majority of these datasets. The reference models (repLLaMA, ColBERTv2) substantially outperform the Hypencoder across nearly all 13 datasets, though they differ considerably in scale and architecture. These results suggest that while the Hypencoder generalizes well to the domains tested in the original work, its zero-shot out-of-domain advantage does not extend uniformly to a broader range of tasks, particularly those involving fact verification (FEVER, Climate-FEVER) and duplicate detection (Quora) where lexical overlap plays a larger role.\looseness=-1

\subsection{Claim 3: Hypencoder Outperforms the Baselines on Hard Tasks}

Table~\ref{tab:hardtasks} reports hard-task results across four conditions: original paper values (Orig.\ Reported), our evaluation of released checkpoints (Orig.\ Ckpts), the base model without fine-tuning (No FT), and our fine-tuned reproduction (FT Ours). TREC DL-Hard uses the base model directly; no fine-tuning applies.

\header{Reproducibility.}
Two issues complicate direct verification. First, the released BE-Base checkpoints produce substantially higher FollowIR scores than reported (e.g., News~'21 nDCG@5: 0.415 vs.\ 0.239), while TOT and DL-Hard scores reproduce accurately. Second, the original Hypencoder fine-tuned checkpoints were incompatible with the released codebase due to mismatched architecture definitions, preventing direct evaluation.

\header{Claim evaluation.}
We assess the claim based on the \emph{relative comparison} between Hypencoder and BE-Base in our own experiments. On  DL-Hard, the Hypencoder outperforms BE-Base (nDCG@10: 0.630 vs.\ 0.601), consistent with the original finding. On FollowIR, our fine-tuned Hypencoder outperforms BE-Base on all three subsets (e.g., Robust~'04 AP: 0.262 vs.\ 0.232; News~'21 nDCG@5: 0.419 vs.\ 0.399). Although absolute scores are inflated for \emph{both} models relative to the original report, the relative ordering is preserved. On TREC TOT, however, the claim is not supported: our fine-tuned Hypencoder underperforms BE-Base (nDCG@10: 0.068 vs.\ 0.077; nDCG@1000: 0.113 vs.\ 0.320), reversing the originally reported ordering. Therefore, we \textbf{partially confirm Claim~3}: the Hypencoder's advantage holds on DL-Hard and FollowIR but not on TREC TOT. The original paper's stronger claim that the performance gap \emph{widens} on harder tasks cannot be fully assessed due to substantial absolute differences between our results and those reported.

\begin{table*}[t]
\caption{Query latency (ms) and nDCG@10 on TREC DL '19/'20. Columns show: Original (BF16, top-$k{=}100$), Reproduced (BF16, top-$k{=}100$), Additional (BF16, top-$k{=}1000$), and Additional (FP32, top-$k{=}1000$). All runs use a single NVIDIA H100 GPU. (–) indicates not applicable.}
\label{tab:reprod_latency}
\centering
\setlength{\tabcolsep}{4pt}
\renewcommand\arraystretch{0.9}
\begin{tabular}{l ccc | ccc | ccc | ccc}
\toprule
& \multicolumn{3}{c}{\makecell{\textbf{Original} \\ (BF16, top-$k=100$)}}
& \multicolumn{3}{c}{\makecell{\textbf{Reproduced} \\ (BF16, top-$k=100$)}}
& \multicolumn{3}{c}{\makecell{\textbf{Additional} \\ (BF16, top-$k=1000$)}}
& \multicolumn{3}{c}{\makecell{\textbf{Additional} \\ (FP32, top-$k=1000$)}} \\
\cmidrule(lr){2-4} \cmidrule(lr){5-7} \cmidrule(lr){8-10} \cmidrule(lr){11-13}
\textbf{Search type or model} & Latency & DL '19 & DL '20 & Latency & DL '19 & DL '20 & Latency & DL '19 & DL '20 & Latency & DL '19 & DL '20 \\
\midrule
Exhaustive            & 1769.8 & 0.742 & 0.731 & 770.6 & 0.742 & 0.731 & 783.7 & 0.742 & 0.731 & 4660.9 & 0.742 & 0.730 \\
Efficient 1 (latency) & 59.6   & 0.702 & 0.730 & 78.4  & 0.698 & 0.718 & 395.9 & 0.716 & 0.731 & 365.0  & 0.708 & 0.730 \\
Efficient 2 (quality) & 231.1  & 0.722 & 0.731 & 238.0 & 0.722 & 0.731 & 671.9 & 0.721 & 0.731 & 667.4  & 0.734 & 0.730 \\
BE-Base with Faiss      & --     & --    & --    & --    & --    & --    & --    & --    & --    & 131.1  & 0.711 & 0.713 \\
\bottomrule
\end{tabular}
\end{table*}

 
\subsection{Claim 4: The Efficient Search Algorithm Improves Query Latency While Retaining Performance}
\label{reprod_claim4}

Table~\ref{tab:reprod_latency} reports query latency and retrieval scores. The two leftmost column groups reproduce the original settings (BF16, top-$k = 100$); the right columns extend this to top-$k = 1{,}000$ and FP32 precision. We note that the use of top-$k = 100$ was undocumented in the original paper but confirmed through author correspondence.

\header{Reproducibility.}
For \textbf{Efficient~2}, we reproduce both latency and retrieval scores nearly exactly. For \textbf{Exhaustive} search, we reproduce the retrieval scores exactly. Our measured latency (770.6\,ms) is substantially lower than the reported value (1769.8\,ms), which is attributable to our use of better hardware (NVIDIA H100 vs. L40S). Our \textbf{Efficient~1} reproduction exhibits slightly lower scores and higher latency than reported; the cause of this minor discrepancy remains unclear.

\header{Claim evaluation.}
Both efficient configurations provide substantial latency reductions with limited loss in retrieval quality. Under the original settings (BF16, top-$k$=100), Efficient~1 achieves a ${\sim}10\times$ speedup over our reproduced exhaustive search (78.4\,ms vs.\ 770.6\,ms) with an nDCG@10 drop of 0.044 on DL~'19 (0.698 vs.\ 0.742). Efficient~2 offers a more conservative trade-off, achieving a ${\sim}3.2\times$ speedup (238.0\,ms vs.\ 770.6\,ms) while largely preserving effectiveness (DL~'19: 0.722 vs.\ 0.742; DL~'20 unchanged at 0.731). Even under top-$k=1000$, both efficient variants remain reasonably close to exhaustive quality, with Efficient~1 reaching 0.716 and Efficient~2 reaching 0.721 on DL~'19, compared to 0.742 for exhaustive search. Among the two, Efficient~2 provides the better quality-preserving trade-off while still reducing latency (671.9\,ms vs.\ 783.7\,ms). Consequently, we \textbf{confirm Claim~4}.

\section{Extension Results} \label{sec:ext_results}

Beyond reproduction, we further analyze the Hypencoder along three dimensions: alternative encoder backbones, query latency relative to a Faiss-based bi-encoder pipeline, and adversarial robustness under query perturbations.

\subsection{Alternative Encoders}

\begin{table}[t]
\centering
\caption{Effect of integrating Hypencoder scoring with alternative encoders on TREC-DL~'19~\&~'20. Standalone model names (e.g., TAS-B) denote the original bi-encoder with inner-product scoring.  \textit{Frozen + FT Hype}: encoder weights frozen, only hyperhead   fine-tuned. \textit{FT End-to-End}: encoder and hyperhead fine-tuned jointly. }
\label{tab:different_encoders}
\renewcommand\arraystretch{0.9}
\begin{tabular}{lccc}
\toprule
\textbf{Model} & nDCG@10 & MRR & R@1000 \\
\midrule
TAS-B   & 0.700 & 0.938 & 0.793 \\
TAS-B Frozen + FT Hype & 0.700 & 0.931 & \textbf{0.797} \\
TAS-B FT End-to-End & \textbf{0.702} & \textbf{0.961} & 0.784 \\
\midrule
RetroMAE   & 0.694 & 0.945 & 0.762 \\
RetroMAE Frozen + FT Hype & 0.682 & 0.944 & 0.783 \\
RetroMAE FT End-to-End & \textbf{0.712} & \textbf{0.954} & \textbf{0.795} \\
\midrule
Contriever   & 0.615 & 0.863 & 0.784 \\
Contriever Frozen + FT Hype & 0.677 & 0.908 & \textbf{0.792} \\
Contriever FT End-to-End & \textbf{0.686} & \textbf{0.930} & 0.772 \\
\bottomrule
\end{tabular}
\end{table}

\header{Setup.}
Full end-to-end Hypencoder training is computationally expensive.  We therefore investigate whether the hyperhead can be integrated with existing pre-trained encoders through fine-tuning. We evaluate two settings: (i)~\textit{Frozen + FT Hype}, where the pre-trained encoder weights are frozen and only the hyperhead is fine-tuned; and (ii)~\textit{FT End-to-End}, where both the encoder and the hyperhead are fine-tuned jointly. We compare each setting against the original bi-encoder baseline.

\header{Results.}
Table~\ref{tab:different_encoders} presents the results. 
Under the frozen encoder setting, we observe mixed changes in nDCG@10 and MRR, but a consistent increase in R@1000 across all three encoders. This suggests that a learned scoring head can leverage embedding spaces optimized for inner-product retrieval to improve recall-oriented behavior without consistently improving top-rank precision. 
Under end-to-end fine-tuning, nDCG@10 and MRR improve across all encoders, while R@1000 shows no consistent trend. This indicates that allowing the model to reshape the embedding space primarily benefits the top of the ranking, consistent with the original Hypencoder results where precision-oriented gains exceeded recall-based gains. 
Overall, the Hypencoder scoring head can be integrated with existing encoders through fine-tuning, offering a lower-cost alternative to training from scratch, though the benefits vary by encoder and metric.\looseness=-1


\subsection{Query Latency Analysis}\label{ext_query_latency}

\begin{figure}[t]
  \centering
  \includegraphics[width=\columnwidth]{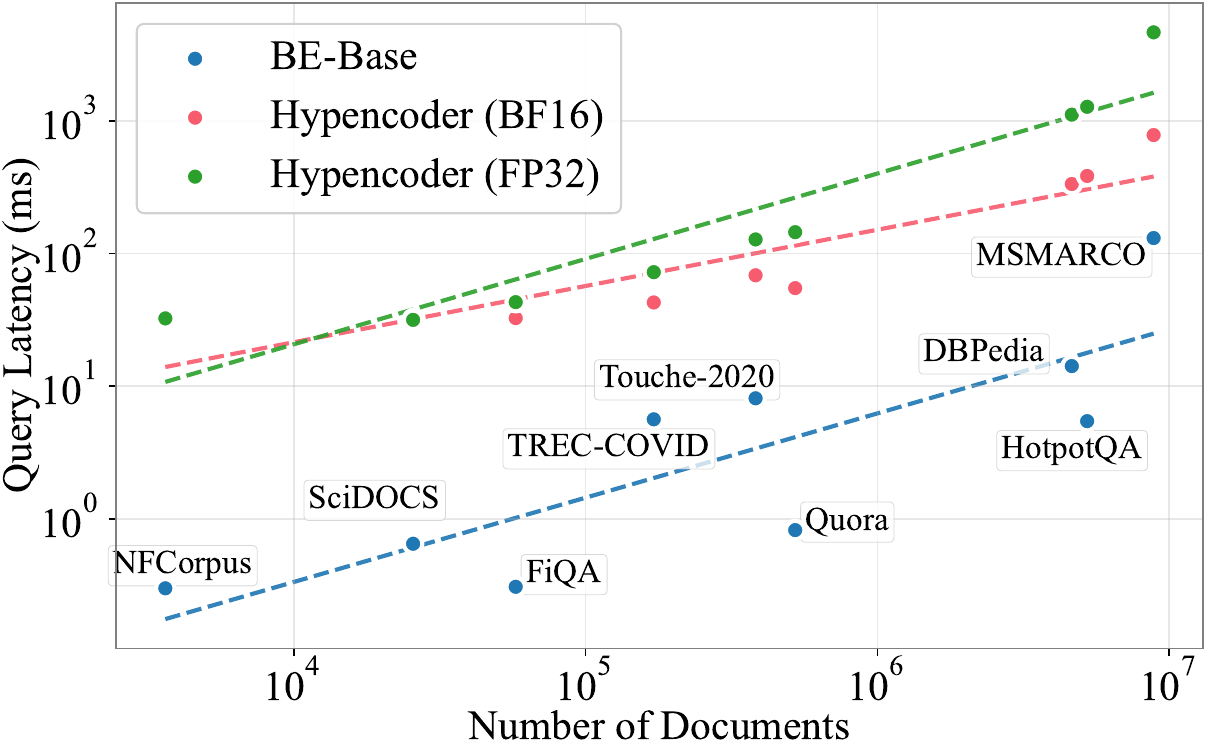}
  \caption{Average per-query latency (ms) vs.\ corpus size (log-log scale) across nine BEIR datasets. BE-Base uses Faiss IndexFlatIP (FP32); Hypencoder uses exhaustive scoring with BF16 and FP32 precision. Dashed lines indicate power-law fits. All experiments on a single NVIDIA H100 GPU.}
  \label{fig:query_latency}
\end{figure}

\begin{figure}[t]
  \centering
  \includegraphics[width=\columnwidth]{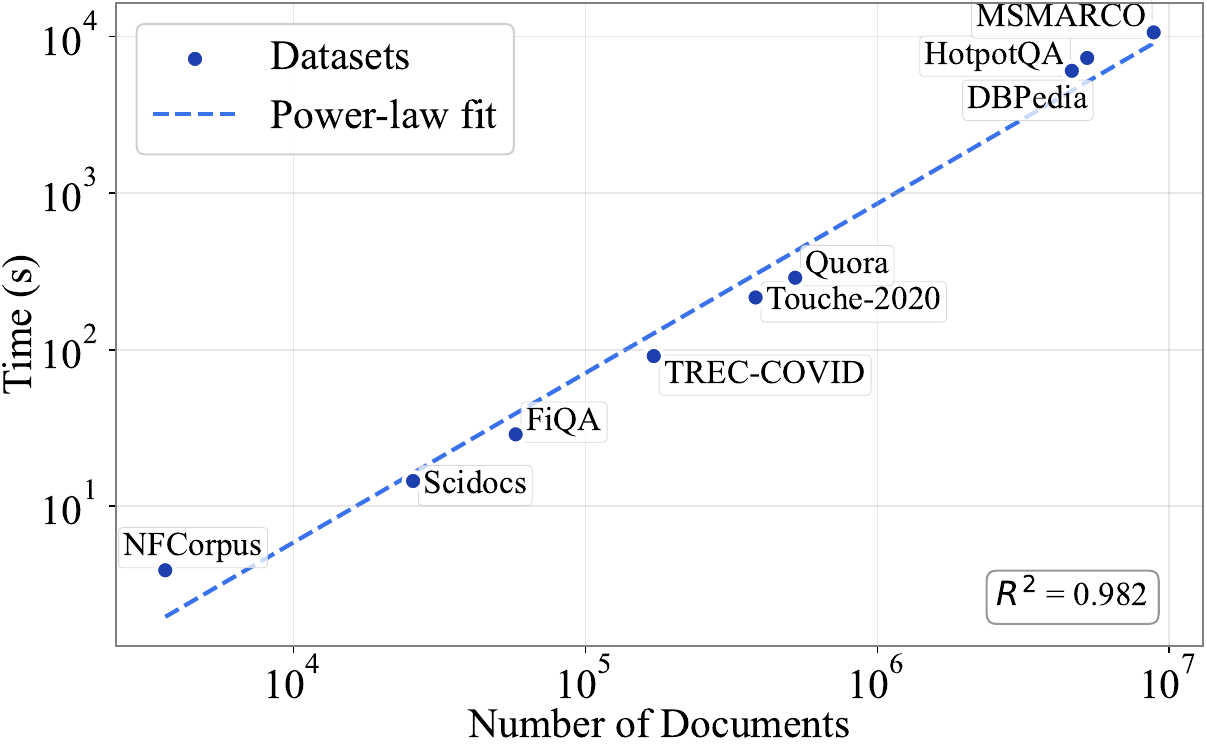}
  \caption{Neighbor graph construction time (seconds) vs.\ corpus size (log-log scale). Dashed lines show a power-law fit. Construction on a single NVIDIA H100 GPU in FP32. }
  \label{fig:graph_construction_time}
\end{figure}

\header{Setup.}
We analyze the query latency of Hypencoder from three perspectives. 
First, we study how top-$k$ and numerical precision affect exhaustive and efficient Hypencoder search. 
Second, we compare Hypencoder against a standard Faiss-based bi-encoder pipeline across datasets with different corpus sizes. 
Third, we measure the offline cost of constructing the neighbor graph required by the efficient search algorithm. 
For the cross-dataset analysis, we select nine BEIR datasets spanning diverse domains and corpus sizes while keeping graph construction tractable.

\header{Effect of Top-$k$ and numerical precision.}
The original paper reports latency with top-$k = 100$ and BF16 inference precision. We extend the evaluation to the more standard top-$k = 1{,}000$ and FP32 precision (right columns of Table~\ref{tab:reprod_latency}). Increasing top-$k$ has a negligible effect on the exhaustive method, as its runtime is dominated by the full corpus scoring rather than the subsequent top-$k$ selection.
For the efficient search algorithm, however, latency increases substantially: a higher top-$k$ lowers the relevance threshold for entering the result queue, causing more documents to be explored and more queue updates. Switching from BF16 to FP32 has a large impact on the exhaustive method (nearly $6\times$ slower), but minimal effect on the efficient search algorithm, whose runtime is largely dominated by graph traversal rather than scoring precision.

\header{Comparison with Faiss.}
To contextualize the Hypencoder's latency, we compare against BE-Base with Faiss (see Section~\ref{sec:metrics} for latency definitions). On MS MARCO (Table~\ref{tab:reprod_latency}, bottom row), Faiss is substantially faster than both the exhaustive Hypencoder and the efficient search algorithm under the top-$k = 1{,}000$ setting.

For the cross-dataset analysis in Figures~\ref{fig:query_latency} and~\ref{fig:graph_construction_time}, we select nine BEIR datasets spanning diverse domains and a wide range of corpus sizes while keeping graph construction tractable. Figure~\ref{fig:query_latency} shows that BE-Base with Faiss is consistently 1--2 orders of magnitude faster than the Hypencoder with FP32 precision.
The gap narrows when BF16 precision is used, particularly for larger corpora: on small datasets like NFCorpus, BF16 and FP32 latencies are nearly identical, but they diverge substantially as corpus size increases. This suggests that lower-bit precision has a proportionally larger effect on latency for large corpora. The relative speedup of Faiss is also larger for datasets with many test queries (e.g., Quora with 10{,}000 queries), as the fixed index-building time is amortized over more queries. Overall, the latency gap reflects the fact that the Hypencoder computes $q$-net scores in batches without leveraging the vectorized search optimizations available in Faiss.

\header{Graph construction overhead.}
Using the efficient search algorithm requires a precomputed neighbor graph. Figure~\ref{fig:graph_construction_time} shows the construction time for several datasets. For MS MARCO, graph construction takes approximately $1.06 \times 10^4$~seconds (${\sim}3$~hours), comparable to the time required for corpus encoding (${\sim}3$~hours with our GPU setup). Although graph construction is a one-time offline cost, it represents significant additional compute compared to the exhaustive Hypencoder method or a Faiss-based bi-encoder pipeline, neither of which requires a neighbor graph. 
For applications requiring repeated querying over a fixed corpus, this one-time cost can be amortized; for frequently updated corpora, however, it may become a practical bottleneck.


\begin{table}[t]
\centering
\caption{Illustrative examples of adversarial query perturbations applied to an original query.}
\label{tab:adversarial_examples}
\setlength{\tabcolsep}{10pt}
\renewcommand\arraystretch{1}
\begin{tabular}{ll}
\toprule
Attack Type & Query Example \\
\midrule
Original Query    & \textit{types of anti depression medication} \\
\midrule
Misspelling & \textit{types of anti depressoin medication} \\
Naturality  & \textit{types of depression medication} \\
Ordering    & \textit{depression of anti types medication} \\
Paraphrasing  & \textit{Types of Antidepressants} \\
Synonymizing     & \textit{kinds of anti depression medication} \\
\bottomrule
\end{tabular}
\vspace{-0.45cm}
\end{table}

\begin{figure}[t]
  \centering
  \includegraphics[width=\columnwidth]{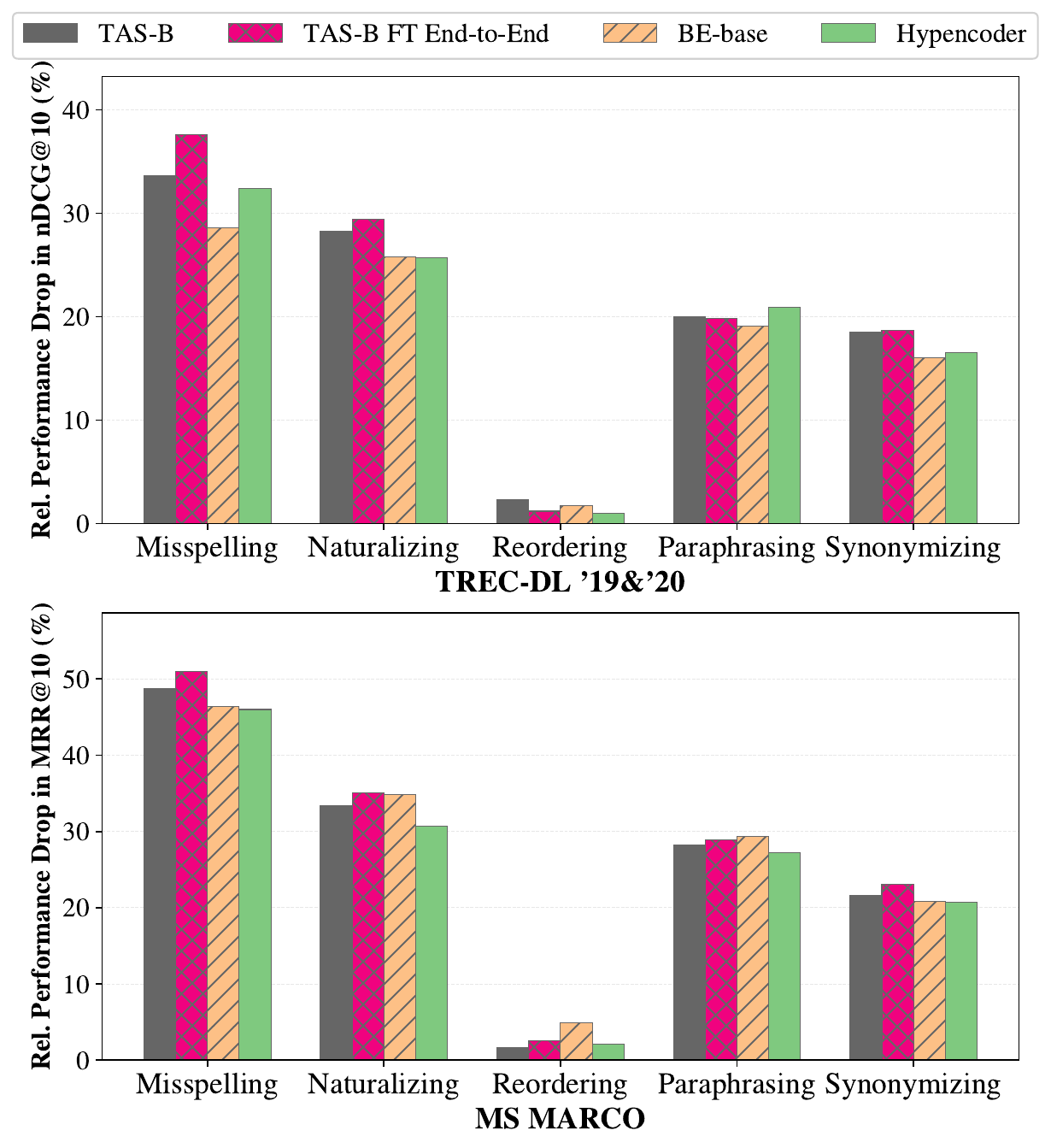}
  \caption{Relative performance drop (\%) under adversarial query perturbations. Top: TREC-DL~'19~\&~'20 (nDCG@10); bottom: MS~MARCO-Dev (MRR@10). Drops are computed relative to each model’s clean-query performance.}
  \label{fig:adversarial}
\end{figure}

\subsection{Adversarial Robustness}

\header{Motivation.}
Prior work has shown that neural retrieval models raise broader concerns beyond effectiveness, including robustness issues and other risks associated with learned text representations~\cite{MorrisKSR23,SeputisLLM25}. 
In the robustness setting, existing work has studied multiple threat models, including corpus-side attacks such as corpus poisoning~\cite{ZhongHWC23_Poisoning,su2024corpus,li2025reproducinghotflip,li2025unsupervised,Li26_ecir} and query-side perturbations~\cite{penha2022evaluating,hagen-etal-2024-revisiting,tasawong-etal-2023-typo,zhuang2021dealing,CharacterBERT}, such as misspellings, paraphrases, and lexical rewrites. 
While both are important, they probe different failure modes. 
We focus on query-side robustness because Hypencoder's main architectural novelty is a query-specific non-linear scorer, the $q$-net, whose generated parameters may be sensitive to small changes in the input query. 
This raises a focused question: do the effectiveness gains of the $q$-net translate into stronger robustness under query perturbations, or does its altered scoring geometry make retrieval more sensitive to such perturbations?

\header{Setup.}
We evaluate whether this affects robustness by comparing two model pairs that isolate the scoring function: (i)~Hypencoder vs.\ BE-Base (different scoring, same training setup), and (ii)~TAS-B \textit{FT End-to-End} vs.\ TAS-B \textit{Base} (same encoder backbone, different scoring).
We apply five perturbation types to the queries following \citet{penha2022evaluating} and \citet{hagen-etal-2024-revisiting}, as illustrated in Table~\ref{tab:adversarial_examples}: \textit{Misspelling} introduces a single typographical error; \textit{Naturality} removes 20\% of query terms (minimum one); \textit{Ordering} performs one random word shuffle; \textit{Paraphrasing} rewrites queries via back-translation through German; and \textit{Synonymizing} replaces one word with a semantically equivalent alternative.

\header{Results.}
Figure~\ref{fig:adversarial} shows the relative performance drops under each perturbation. 
For misspelling attacks on the TREC datasets, inner-product models exhibit smaller performance drops than their Hypencoder counterparts, indicating greater robustness to this type of lexical noise. 
For the remaining attacks, neither scoring function shows a consistent advantage: performance drops are similar and vary by attack, dataset, and metric. 
Overall, $q$-net scoring does not systematically reduce robustness to query perturbations.\looseness=-1

\section{Conclusion}
We presented a reproducibility study of the Hypencoder framework~\cite{killingback2025hypencoder}.
Our findings for each claim are as follows:
\textbf{Claim~1} (in-domain): \emph{confirmed}---Hypencoder consistently outperforms BE-Base and other comparably-scaled baselines on TREC DL and MS MARCO Dev.
\textbf{Claim~2} (out-of-domain): \emph{confirmed within the original scope}---results on the five originally reported BEIR datasets reproduce accurately, though the advantage does not extend uniformly to the broader BEIR benchmark.
\textbf{Claim~3} (hard tasks): \emph{partially confirmed}---Hypencoder outperforms BE-Base on DL-Hard and all three FollowIR subsets, but not on TREC TOT, where the relative ordering reverses. Checkpoint incompatibility and sensitivity to fine-tuning hyperparameters prevent full verification of the originally reported results.
\textbf{Claim~4} (efficient search): \emph{confirmed}---both efficient configurations substantially reduce latency with limited quality loss.
Beyond reproduction, our extension experiments show that the Hypencoder scoring head can be integrated with existing pre-trained encoders through fine-tuning, offering a lower-cost alternative to training from scratch. We also find that the $q$-net's non-linear scoring does not introduce systematic robustness degradation under adversarial query perturbations. However, even with efficient search, Hypencoder remains slower than Faiss-based bi-encoder retrieval, highlighting a practical efficiency trade-off.

Our conclusions should be interpreted in light of one important limitation. We did not train the hypernetwork generating the $q$-net from scratch, and instead relied on the optimized checkpoints released by the original authors. Given the substantial training cost reported in the original paper (six days on two A100 GPUs) and the scope of our reproducibility study, this was a pragmatic choice. This limitation also motivated our extension on integrating the hyperhead with alternative pre-trained encoders. Nevertheless, it remains important for future work to examine whether our findings, as well as those of the original paper, continue to hold when the full Hypencoder is trained from scratch.

Looking ahead, one promising direction is to use $q$-net scoring for re-ranking rather than full first-stage retrieval. Under such a setup, a standard bi-encoder with a Faiss-backed index could first retrieve a candidate set, followed by fine-grained re-ranking with the $q$-net. This design may preserve part of Hypencoder's expressive scoring benefit while reducing the latency gap relative to Faiss-based retrieval and avoiding the cost of constructing a neighbor graph over the full corpus.

\begin{acks}
This work was in part supported by the China Scholarship Council (202308440220) and in part by the Informatics Institute (IvI), University of Amsterdam. We want to thank Julian Killingback, author of the original paper, for his quick and helpful responses to our questions. 
The authors acknowledge the peoples of the Woi Wurrung and Boon Wurrung language groups of the eastern Kulin Nation on whose unceded lands ACM SIGIR 2026 was hosted. We pay our respects to their Elders past and present, and extend that respect to all Aboriginal and Torres Strait Islander peoples today and their continuing connection to land, sea, sky, and community.
\end{acks}


\bibliographystyle{ACM-Reference-Format}
\balance
\bibliography{references}


\end{document}